\documentclass[showpacs,aps,graphicx,twocolumn]{revtex4}%and
\usepackage{graphicx}

\begin{document}

\title{Efficient $N$-particle $W$ state concentration with different parity check gates}

\author{Yu-Bo Sheng,$^{1,4}$\footnote{Email address:
shengyb@njupt.edu.cn} Lan Zhou,$^2$ Yu-Wei Sheng,$^3$
 Sheng-Mei Zhao,$^{1,4}$ }
\address{$^1$ Institute of Signal Processing  Transmission, Nanjing
University of Posts and Telecommunications, Nanjing, 210003,  China\\
 $^2$Beijing National Laboratory for Condensed Matter Physics, Institute of Physics,\\
Chinese Academy of Sciences, Beijing 100190, China\\
$^3$School of Computer Science, Beijing University of Posts and
Telecommunications, Beijing 100876, China\\
 $^4$Key Lab of Broadband Wireless Communication and Sensor Network
 Technology,
 Nanjing University of Posts and Telecommunications, Ministry of
 Education, Nanjing, 210003,
 China\\}

\date{\today }

\begin{abstract}
We present an universal way to  concentrate an arbitrary $N$-particle
less-entangled $W$ state into a maximally entangled $W$ state with
different parity check gates. It comprises two
protocols. The first protocol is based on the linear optical elements say the
 partial parity check gate and
the second one uses the quantum nondemolition (QND) to construct the
complete parity check gate. Both of
which can achieve the concentration task. These protocols have several
advantages. First, it can obtain a maximally entangled W state only
with the help of some single photons, which greatly reduces the number of
entanglement resources. Second, in the first protocol, only linear
optical elements are required which is feasible with current
techniques. Third, in the second protocol, it can be repeated to
perform the concentration step and get a higher success probability.
All these advantages make it be useful in current quantum
communication and computation applications.

\end{abstract}
 \maketitle

\section{Introduction}
Entanglement is the important quantum resource  in both quantum
communication and computation \cite{book,rmp}. The applications of entanglement
information processings
 such as quantum teleportation \cite{teleportation,cteleportation}, quantum key distribution (QKD) \cite{Ekert91,QKDdeng1,QKDdeng2},
 quantum dense coding \cite{densecoding1,densecoding2},  quantum secret sharing \cite{QSS1,QSS2,QSS3} and quantum
 secure direct communication (QSDC) \cite{QSDC1,QSDC2,QSDC3} all resort the entanglement
 for setting up the quantum channel between long distance locations.
Unfortunately, during the practical transmission, an entangled
quantum system can not avoid the channel noise that comes from the
environment, which will degrade the entanglement. It will make a
maximally entangled state system become a mixed one or a partially
entangled one. Therefore, these nonmaximally entangled systems will
decrease the security of a QKD protocol if it is used to set up the
quantum channel. Moreover, they also will decrease the fidelity of
quantum dense coding and quantum teleportation.

Entanglement purification is a powerful tool for parties to improve
the fidelity of the entangled state from a mixed entangled
ensembles \cite{C.H.Bennett1,D. Deutsch,Pan1,Pan2,M.
Murao,shengpra,M. Horodecki,Yong,lixhepp,dengonestep1,wangc1,Simon,sangouard}. On
the other hand, the entanglement concentration protocol (ECP) is focused
on the pure less-entangled system, which can be used to recover a
pure less-entangled state into a pure maximally entangled state with only
local operation and classical
communications \cite{C.H.Bennett2,swapping1,swapping2,Yamamoto1,zhao1,wangxb,shengpra2,shengqic,
shengsinglephotonconcentration,shengWthree,cao,zhanglihua,wang,yildiz,dengsingle}.
Most of the ECPs such as the Schmidt decomposition protocol proposed by Bennett
\emph{et al.} \cite{C.H.Bennett2}, the ECPs based on entanglement
swapping \cite{swapping1,swapping2}, linear
optics \cite{zhao1,wangxb,Yamamoto1}, and cross-Kerr
nonlinearity \cite{shengpra2,shengsinglephotonconcentration} are all focused on the Bell states and
multi-partite Greenberger-Horne-Zeilinger (GHZ) states. Because all
the ECPs for Bell stats can be easily extended to the GHZ states.

On the other hand, the $W$ state, which has the different entanglement
structure and can not be convert to the GHZ state directly with only
local operation and classical communication, has began to receive
attention both in theory and experiment \cite{W1,W2,W3,W4,W5,W6}.  Agrawal and Pati presented a perfect
teleportation and superdense coding with $W$ states in 2006 \cite{W2}. In 2010, Tamaryan \emph{et al.} discussed
the universal behavior of the geometric entanglement measure of many-qubit $W$ states \cite{W4}. Eibl \emph{et al.} also realized a three-qubit entangled $W$ state in experiment \cite{W5}. Several ECPs for less-entangled  $W$ state were also proposed \cite{cao,zhanglihua,wang,yildiz,shengWthree}.
In 2003, Cao and Yang has discussed the $W$ state
concentration with the help of joint unitary
transformation\cite{cao}. In 2007, a $W$ state ECP
 based on the Bell-state measurement has been
proposed \cite{zhanglihua}. Then in 2010, Wang \emph{et al.} have
proposed an ECP which focuses on a special kind of W
state \cite{wang}. Recently, Yildiz proposed  an optimal distillation
of three-qubit asymmetric $W$ states \cite{yildiz}. We also have
proposed an ECP with both linear optics and cross-Kerr for
three-particle $W$ state \cite{shengWthree}. Unfortunately, these ECPs described above all
 focus on the three-particle $W$ state and they are mostly to concentrate some  $W$ states with  the special structures.

In this paper, we will present an ECP for arbitrary multi-partite
polarized $W$ entangled systems. We will describe this protocol in
two different ways. First,  we use the partial parity check (PPC)
gate constructed by linear optics to perform this protocol. Second,
we introduce the complete parity check (CPC) gate to achieve this
task. Compared with other conventional ECPs, we only resort the
single photon as an auxiliary which largely reduce the consumed
quantum resources. Moreover, with the help of CPC gate, this
protocol can be repeated and get a higher success probability. This
paper is organized as follows: in Sec. II, we first briefly explain
our PPC gate and CPC gate. in Sec. III, we describe our ECP with
both PPC and CPC gates respectively. In Sec. IV, we make a
discussion and summary.

\section{Parity check gate}
Before we start to explain this protocol, we first briefly describe
the parity check gate. Parity check gate is the basic element in
quantum communication and computation. It can be used to construct
the controlled-not (CNOT) gate \cite{pittman,QND1}. It also
can be used to perform the entanglement purification \cite{Pan1,Pan2} and
concentration protocol \cite{zhao1,Yamamoto1}.

\subsection{partial parity check gate}
There are two different kinds of parity check gates.  One is the
partial parity check (PPC) gate and the other is the complete parity check (CPC)  gate  .
In optical system, a polarization beam
splitter (PBS) is essentially  a good candidate for PPC gate as shown in Fig.1. Suppose that two polarized photons
of the form
\begin{eqnarray}
|\varphi_{1}\rangle=\alpha|H\rangle+\beta|V\rangle,
|\varphi_{2}\rangle=\gamma|H\rangle+\delta|V\rangle,
\end{eqnarray}
entrance into the PBS from different spatial modes.
Here $|\alpha|^{2}+|\beta|^{2}=1$, and $|\gamma|^{2}+|\delta|^{2}=1$.
$|H\rangle$ and $|V\rangle$ represent the horizonal and the vertical
polarization of the photons, respectively.

\begin{figure}[!h]%[tpb]
\begin{center}
\includegraphics[width=6cm,angle=0]{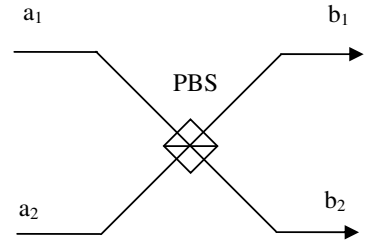}
\caption{A schematic drawing of our PPC gate. It is constructed by a
polarization beam splitter(PBS). It is used to transfer a
$|H\rangle$ polarization photon and to reflect a $|V\rangle$
polarization photon. }
\end{center}
\end{figure}

Let $|\varphi_{1}\rangle$ be in the spatial mode $a_{1}$ and
 $|\varphi_{2}\rangle$ be in the spatial mode $a_{2}$. The whole
 system can be described as
\begin{eqnarray}
|\varphi_{1}\rangle&\otimes&|\varphi_{2}\rangle=(\alpha|H\rangle_{a_{1}}+\beta|V\rangle_{a_{1}})\otimes(\gamma|H\rangle_{a_{2}}+\delta|V\rangle_{a_{2}})\nonumber\\
&=&\alpha\gamma|H\rangle_{a_{1}}|H\rangle_{a_{2}}+\beta\delta|V\rangle_{a_{1}}|V\rangle_{a_{2}}\nonumber\\
&+&\alpha\delta|H\rangle_{a_{1}}|V\rangle_{a_{2}}+\beta\gamma|V\rangle_{a_{1}}|H\rangle_{a_{2}}
\end{eqnarray}
Then after passing through the PBS, it evolves as
\begin{eqnarray}
&\rightarrow&\alpha\gamma|H\rangle_{b_{1}}|H\rangle_{b_{1}}+\beta\delta|V\rangle_{b_{1}}|V\rangle_{b_{2}}\nonumber\\
&+&\alpha\delta|H\rangle_{b_{1}}|V\rangle_{b_{1}}+\beta\gamma|V\rangle_{b_{2}}|H\rangle_{b_{2}}.
\end{eqnarray}
From above description, items $|H\rangle_{b_{1}}|H\rangle_{b_{1}}$
and $|V\rangle_{b_{1}}|V\rangle_{b_{2}}$, say the even parity states
will lead the output modes $b_{1}$ and $b_{2}$ both exactly contain
only one photon. But items $|H\rangle_{b_{1}}|V\rangle_{b_{1}}$ and
$|V\rangle_{b_{2}}|H\rangle_{b_{2}}$ will lead the two photons be in
the same output mode, which cannot be distinguished. Based on the
post selection principle, only the even parity state is the
successful case. Therefore, the total success probability is
$|\alpha\gamma|^{2}+|\beta\delta|^{2}<1$. This is the reason that we
call it PPC gate.

\subsection{complete parity check gate}
Another parity check gate say CPC gate is shown in Fig. 2.  We adopt the
cross-Kerr nonlinearity to construct the CPC gate. Cross-Kerr
nonlinearity has been widely used in quantum information
processing\cite{{lin1,he1,shengbellstateanalysis,qi}}. In general,
the Hamiltonian of the cross-Kerr nonlinearity is described as
$H=\hbar\chi \hat{n_{a}}\hat{n_{b}}$, where the $\hbar\chi$ is the
coupling strength of the nonlinearity. It is decided by the material
of cross-Kerr. The $ \hat{n_{a}}(\hat{n_{b}})$ are the number
operator for mode $a(b)$\cite{QND1,QND2}.

Now we reconsider the two photon system
$|\varphi_{1}\rangle\otimes|\varphi_{2}\rangle$ coupled with the
coherent state $|\alpha\rangle$. From Fig. 2,  the whole system evolves as
\begin{eqnarray}
|\varphi_{1}\rangle&\otimes|&\varphi_{2}\rangle\otimes|\alpha\rangle=(\alpha\gamma|H\rangle_{a_{1}}|H\rangle_{a_{2}}+\beta\delta|V\rangle_{a_{1}}|V\rangle_{a_{2}}\nonumber\\
&+&\alpha\delta|H\rangle_{a_{1}}|V\rangle_{a_{2}}+\beta\gamma|V\rangle_{a_{1}}|H\rangle_{a_{2}})\otimes|\alpha\rangle\nonumber\\
&\rightarrow&(\alpha\gamma|H\rangle_{b_{1}}|H\rangle_{b_{1}}+\beta\delta|V\rangle_{b_{1}}|V\rangle_{b_{2}})|\alpha\rangle\nonumber\\
&+&\alpha\delta|H\rangle_{b_{1}}|V\rangle_{b_{1}}|\alpha
e^{-i2\theta}\rangle+\beta\gamma|V\rangle_{b_{2}}|H\rangle_{b_{2}}|\alpha
e^{i2\theta}\rangle.\nonumber\\
\end{eqnarray}
\begin{figure}[!h]%[tpb]
\begin{center}
\includegraphics[width=7cm,angle=0]{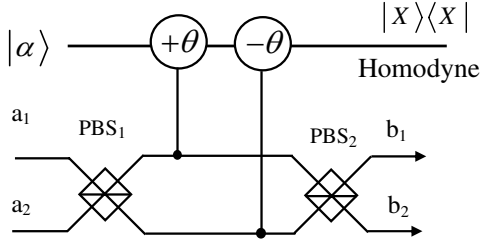}
\caption{A schematic drawing of our CPC gate. }
\end{center}
\end{figure}
It is obvious to see that the even parity states make the coherent
state $|\alpha\rangle$ pick up no phase shift, but the odd parity
state $|H\rangle_{b_{1}}|V\rangle_{b_{2}}$ makes the coherent state
pick up the phase shift $-2\theta$. The other odd parity state
$|V\rangle_{b_{1}}|H\rangle_{b_{2}}$ make the coherent state pick up
the phase shift with $2\theta$. With a general homodyne-heterodyne
measurement, the phase shift $2\theta$ and $-2\theta$ can not be
distinguished\cite{QND1}. Then one can distinguish the different parity state
according to their different phase shifts. So the success probability
of the initial states collapsing to the even and odd parity state is
$|\alpha\gamma|^{2}+|\beta\delta|^{2}+|\alpha\delta|^{2}+|\beta\gamma|^{2}=1$,
in principle. So we call it CPC gate.
  Compared with the PPC gate, the success probability for CPC gate can reach the max value 1 but
  the PPC gate cannot reach 1. Another advantage of
the CPC gate is that we get the both even and odd parity state by
measuring the phase shift of the coherent state. That is to say, we
do not need to measure the two photons directly. So after the
measurement, the two photons can be remained. It is so called
quantum nondemolition(QND) measurement. But in PPC gate, we should use the post
selection principle  to detect the two photons being in the different spatial
modes by coincidence counting. After both detectors register the photons with a success case,
the photons are destroyed and cannot be
used further more.

\section{$N$-particle less-entangled $W$ state concentration with parity check gate}
\subsection{$N$-particle less-entangled $W$ state concentration with PPC gate}
In this section, we start to describe our $N$-particle ECP with
PPC gate. An $N$-particle $W$ state can be described as
\begin{eqnarray}
|\Psi\rangle_{N}&=&\alpha_{1}|V\rangle_{1}|H\rangle_{2}|H\rangle_{3}+\cdots+|H\rangle_{N-1}|H\rangle_{N}\nonumber\\
&+&\alpha_{2}|H\rangle_{1}|V\rangle_{2}|H\rangle_{3}+\cdots+|H\rangle_{N-1}|H\rangle_{N}\nonumber\\
&+&\cdots+\alpha_{N}|H\rangle_{1}|H\rangle_{2}+\cdots+|H\rangle_{N-1}|V\rangle_{N}\nonumber\\
&=&\alpha_{1}|V\rangle_{1}|H\rangle_{2}|\widetilde{H}\rangle^{N-2}+\alpha_{2}|H\rangle_{1}|V\rangle_{2}|\widetilde{H}\rangle^{N-2}\nonumber\\
&+&\cdots+\alpha_{N}|H\rangle_{1}|H\rangle_{2}|\widetilde{H}\rangle^{N-3}|V\rangle_{N}.\label{partial
W state}
\end{eqnarray}
where $|\alpha_{1}|^{2}+|\alpha_{2}|^{2}+\cdots+|\alpha_{N}|^{2}=1$.
In order to explain this ECP clearly simply, we let $\alpha_{1}$,
$\alpha_{2}$, $\cdots$ be real.  Certainly, this ECP is also suitable for the case
of  $\alpha_{1}$,
$\alpha_{2}$, $\cdots$ being complex. $|\widetilde{H}\rangle^{N-2}$ means
that the $N-2$ photons say $|H\rangle_{3}|H\rangle_{4}\cdots|H\rangle_{N}$ are all in the $|H\rangle$ polarization.

\begin{figure}[!h]%[tpb]
\begin{center}
\includegraphics[width=9cm,angle=0]{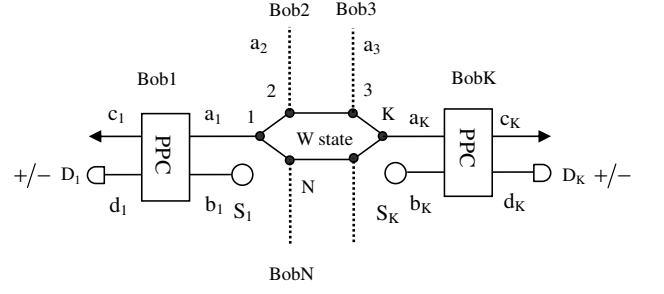}
\caption{A schematic drawing of our ECP with PPC
gate. Each parties except Bob2 own the PPC gate and perform
the parity check. If they pick up the even parity state, it is successful, otherwise, it is a failure.}
\end{center}
\end{figure}

From Fig. 3, the $N$-photon less-entangled $W$ state  of the form Eq.(\ref{partial
W state}) is distributed to $N$ parties, saies Bob1, Bob2, $\cdots$, Bob$N$. Bob1
receives the photon of number $1$ in the spatial mode $a_{1}$.
Bob2 receives the number $2$ in the spatial mode $a_{2}$, and Bob$N$
receives the photon number $N$  in the spatial mode $a_{N}$. That is to say, each of the parties
owns one photon.

The  principle of our ECP with PPC gate is shown in Fig. 3. The basic
idea of realizing  the concentration is to use  the local operation and classical
communication to make each coefficients on each items of Eq. (\ref{partial
W state})  all equal to $\alpha_{2}$. If
all coefficients are equal, they can be regarded as a common factor
and can be neglected. The remained state is essentially the maximally
entangled $W$ state. Thus, the whole process can be briefly described
as follows: we first divide the whole procedure into $N-1$ steps. In
each step, each party say Bob$K$ should first prepare a single photon. In Fig. 3, the single-photon
sources $S_{1}$, $S_{3}$, $\cdots$, $S_{K}$, $\cdots$, $S_{N}$ are used to prepare the single photons locally.
Then he performs a parity check measurement for his two photons. The one comes
from the single photon he  prepared, and the other is the photon
from the less-entangled  $W$ state. If the parity check measurement is successful, then
he asks the other to perform the further operation.

Bob1  first perform the parity check  on the photon of number 1
and the prepared single photon. The single-photon resource $S_{1}$ for Bob1 prepares a single photon
in the spatial mode $b_{1}$ of the form
\begin{eqnarray}
|\Phi\rangle_{1}=\frac{\alpha_{1}}{\sqrt{\alpha_{1}^{2}+\alpha_{2}^{2}}}|H\rangle
+\frac{\alpha_{2}}{\sqrt{\alpha_{1}^{2}+\alpha_{2}^{2}}}|V\rangle.\label{S1}
\end{eqnarray}
Then the initial less-entangled $W$ state $|\Psi\rangle_{N}$ combined with $|\Phi\rangle_{1}$
can be described as
\begin{eqnarray}
&&|\Psi\rangle_{N+1}=|\Psi\rangle_{N}\otimes|\Phi\rangle_{1}
=(\alpha_{1}|V\rangle_{1}|H\rangle_{2}|\widetilde{H}\rangle^{N-2}\nonumber\\
&+&\alpha_{2}|H\rangle_{1}|V\rangle_{2}|\widetilde{H}\rangle^{N-2}
+\cdots
+\alpha_{N}|H\rangle_{1}|H\rangle_{2}|\widetilde{H}\rangle^{N-3}|V\rangle_{N})\nonumber\\
&\otimes&(\frac{\alpha_{1}}{\sqrt{\alpha_{1}^{2}+\alpha_{2}^{2}}}|H\rangle
+\frac{\alpha_{2}}{\sqrt{\alpha_{1}^{2}+\alpha_{2}^{2}}}|V\rangle)\nonumber\\
&=&\frac{\alpha_{1}^{2}}{\sqrt{\alpha_{1}^{2}+\alpha_{2}^{2}}}|H\rangle|V\rangle_{1}|H\rangle_{2}|\widetilde{H}\rangle^{N-2}\nonumber\\
&+&\frac{\alpha_{2}^{2}}{\sqrt{\alpha_{1}^{2}+\alpha_{2}^{2}}}|V\rangle|H\rangle_{1}|V\rangle_{2}|\widetilde{H}\rangle^{N-2}\nonumber\\
&+&\frac{\alpha_{1}\alpha_{2}}{\sqrt{\alpha_{1}^{2}+\alpha_{2}^{2}}}|H\rangle|H\rangle_{1}|V\rangle_{2}|\widetilde{H}\rangle^{N-2}\nonumber\\
&+&\frac{\alpha_{1}\alpha_{3}}{\sqrt{\alpha_{1}^{2}+\alpha_{2}^{2}}}|H\rangle|H\rangle_{1}|H\rangle_{2}|V\rangle_{3}|\widetilde{H}\rangle^{N-3}\nonumber\\
&+&\cdots\nonumber\\
&+&\frac{\alpha_{1}\alpha_{N}}{\sqrt{\alpha_{1}^{2}+\alpha_{2}^{2}}}|H\rangle|H\rangle_{1}|H\rangle_{2}|\widetilde{H}\rangle^{N-3}|V\rangle_{N}\nonumber\\
&+&\frac{\alpha_{1}\alpha_{2}}{\sqrt{\alpha_{1}^{2}+\alpha_{2}^{2}}}|V\rangle|V\rangle_{1}|H\rangle_{2}|\widetilde{H}\rangle^{N-2}\nonumber\\
&+&\frac{\alpha_{2}\alpha_{3}}{\sqrt{\alpha_{1}^{2}+\alpha_{2}^{2}}}|V\rangle|H\rangle_{1}|H\rangle_{2}|V\rangle_{3}|\widetilde{H}\rangle^{N-3}\nonumber\\
&+&\cdots\nonumber\\
&+&\frac{\alpha_{2}\alpha_{N}}{\sqrt{\alpha_{1}^{2}+\alpha_{2}^{2}}}|V\rangle|H\rangle_{1}|H\rangle_{2}|\widetilde{H}\rangle^{N-3}|V\rangle_{N}.
\end{eqnarray}
After passing through the PPC gate in Bob1's location, Bob1 only picks up the even
parity state in the spatial mode $c_{1}$ and $d_{1}$. Therefore, the
above state collapses to
\begin{eqnarray}
|\Psi\rangle'_{N+1}&=&\frac{\alpha_{1}\alpha_{2}}{\sqrt{\alpha_{1}^{2}+\alpha_{2}^{2}}}|H\rangle|H\rangle_{1}|V\rangle_{2}|\widetilde{H}\rangle^{N-2}\nonumber\\
&+&\frac{\alpha_{1}\alpha_{3}}{\sqrt{\alpha_{1}^{2}+\alpha_{2}^{2}}}|H\rangle|H\rangle_{1}|H\rangle_{2}|V\rangle_{3}|\widetilde{H}\rangle^{N-3}\nonumber\\
&+&\cdots\nonumber\\
&+&\frac{\alpha_{1}\alpha_{N}}{\sqrt{\alpha_{1}^{2}+\alpha_{2}^{2}}}|H\rangle|H\rangle_{1}|H\rangle_{2}|\widetilde{H}\rangle^{N-3}|V\rangle_{N}\nonumber\\
&+&\frac{\alpha_{1}\alpha_{2}}{\sqrt{\alpha_{1}^{2}+\alpha_{2}^{2}}}|V\rangle|V\rangle_{1}|H\rangle_{2}|\widetilde{H}\rangle^{N-2}.\label{collapse11}
\end{eqnarray}
It can be rewritten as
\begin{eqnarray}
&&|\Psi\rangle''_{N+1}=\frac{\alpha_{2}}{\sqrt{2\alpha_{2}^{2}+\alpha_{3}^{2}+\cdots+\alpha_{N}^{2}}}|H\rangle|H\rangle_{1}|V\rangle_{2}|\widetilde{H}\rangle^{N-2}\nonumber\\
&+&\frac{\alpha_{2}}{\sqrt{2\alpha_{2}^{2}+\alpha_{3}^{2}+\cdots+\alpha_{N}^{2}}}|V\rangle|V\rangle_{1}|V\rangle_{2}|\widetilde{H}\rangle^{N-2}\nonumber\\
&+&\frac{\alpha_{3}}{\sqrt{2\alpha_{2}^{2}+\alpha_{3}^{2}+\cdots+\alpha_{N}^{2}}}|H\rangle|H\rangle_{1}|H\rangle_{2}|V\rangle_{3}|\widetilde{H}\rangle^{N-3}\nonumber\\
&+&\cdots\nonumber\\
&+&\frac{\alpha_{N}}{\sqrt{2\alpha_{2}^{2}+\alpha_{3}^{2}+\cdots+\alpha_{N}^{2}}}|H\rangle|H\rangle_{1}|\widetilde{H}\rangle^{N-2}|V\rangle_{N}\label{rewrittecollapse11}.
\end{eqnarray}
Finally, Bob1 measures the photon in the spatial mode $d_{1}$ (the first photon in Eq. (\ref{rewrittecollapse11})) in the
basis $|\pm\rangle$, with
$|\pm\rangle=\frac{1}{\sqrt{2}}(|H\rangle\pm|V\rangle)$. Then they will get
\begin{eqnarray}
|\Psi^{\pm}\rangle^{1}_{N}&=&\pm\frac{\alpha_{2}}{\sqrt{2\alpha_{2}^{2}+\alpha_{3}^{2}+\cdots+\alpha_{N}^{2}}}|V\rangle_{1}|H\rangle_{2}|\widetilde{H}\rangle^{N-2}\nonumber\\
&+&\frac{\alpha_{2}}{\sqrt{2\alpha_{2}^{2}+\alpha_{3}^{2}+\cdots+\alpha_{N}^{2}}}|H\rangle_{1}|V\rangle_{2}|\widetilde{H}\rangle^{N-2}\nonumber\\
&+&\frac{\alpha_{3}}{\sqrt{2\alpha_{2}^{2}+\alpha_{3}^{2}+\cdots+\alpha_{N}^{2}}}|H\rangle_{1}|H\rangle_{2}|V\rangle_{3}|\widetilde{H}\rangle^{N-3}\nonumber\\
&+&\cdots\nonumber\\
&+&\frac{\alpha_{N}}{\sqrt{2\alpha_{2}^{2}+\alpha_{3}^{2}+\cdots+\alpha_{N}^{2}}}|H\rangle_{1}|\widetilde{H}\rangle^{N-2}|V\rangle_{N}.\nonumber\\\label{W11}
\end{eqnarray}
The superscription 1 means that they perform the concentration on
the first particle. If the measurement result is $|+\rangle$, they
will get $|\Psi^{+}\rangle^{1}_{N}$. If the result is  $|-\rangle$,
they will get $|\Psi^{-}\rangle^{1}_{N}$. In order to get $|\Psi^{+}\rangle^{1}_{N}$, one of the parties, Bob$1$, Bob$2$,
$\cdots$ should perform a local operation of phase rotation on
 his photon. The total success
probability is
\begin{eqnarray}
P^{1}&=&\frac{2\alpha^{2}_{1}\alpha^{2}_{2}+\alpha^{2}_{1}(\alpha^{2}_{3}+\alpha^{2}_{4}+\cdots+\alpha^{2}_{N})}{\alpha^{2}_{1}+\alpha^{2}_{2}}\nonumber\\
&=&\frac{\alpha^{2}_{1}(2\alpha^{2}_{2}+\alpha^{2}_{3}+\alpha^{2}_{4}+\cdots+\alpha^{2}_{N})}{\alpha^{2}_{1}+\alpha^{2}_{2}}.
\end{eqnarray}
Compared with Eq.(\ref{partial W state}), the coefficient of
$\alpha_{1}$ has disappeared in  the state of Eq.(\ref{W11}).

The next step is to
prepare another single photon in single-photon source $S_{3}$ in the
spatial mode $b_{3}$  of the form
\begin{eqnarray}
|\Phi\rangle_{3}=\frac{\alpha_{2}}{\sqrt{\alpha_{2}^{2}+\alpha_{3}^{2}}}|V\rangle
+\frac{\alpha_{3}}{\sqrt{\alpha_{2}^{2}+\alpha_{3}^{2}}}|H\rangle.\label{S3}
\end{eqnarray}

Following  the same principle described above, Bob3 lets the photon
of number 3 in $|\Psi^{+}\rangle^{1}_{N}$ in the spatial mode
$a_{3}$ combined with the single photon $|\Phi\rangle_{3}$ in the
spatial mode $b_{3}$ pass through his PPC gate. Then the whole
system  evolves to
\begin{widetext}
\begin{eqnarray}
|\Psi^{+}\rangle^{1}_{N}\otimes|\Phi\rangle_{3}
&\rightarrow&\frac{\alpha_{2}\alpha_{3}}{\sqrt{2\alpha_{2}^{2}+\alpha_{3}^{2}+\cdots+\alpha_{N}^{2}}\sqrt{\alpha_{2}^{2}+\alpha_{3}^{2}}}|V\rangle_{1}|H\rangle_{2}|H\rangle_{3}|H\rangle|\widetilde{H}\rangle^{N-3}\nonumber\\
&+&\frac{\alpha_{2}\alpha_{3}}{\sqrt{2\alpha_{2}^{2}+\alpha_{3}^{2}+\cdots+\alpha_{N}^{2}}\sqrt{\alpha_{2}^{2}+\alpha_{3}^{2}}}|H\rangle_{1}|V\rangle_{2}|H\rangle_{3}|H\rangle|\widetilde{H}\rangle^{N-3}\nonumber\\
&+&\frac{\alpha_{2}\alpha_{3}}{\sqrt{2\alpha_{2}^{2}+\alpha_{3}^{2}+\cdots+\alpha_{N}^{2}}\sqrt{\alpha_{2}^{2}+\alpha_{3}^{2}}}|H\rangle_{1}|H\rangle_{2}|V\rangle_{3}|V\rangle|\widetilde{H}\rangle^{N-3}\nonumber\\
&+&\cdots\nonumber\\
&+&\frac{\alpha_{3}\alpha_{N}}{\sqrt{2\alpha_{2}^{2}+\alpha_{3}^{2}+\cdots+\alpha_{N}^{2}}\sqrt{\alpha_{2}^{2}+\alpha_{3}^{2}}}|H\rangle_{1}|H\rangle_{2}|H\rangle_{3}|H\rangle|\widetilde{H}\rangle^{N-3}|V\rangle_{N},
\end{eqnarray}
\end{widetext}
If he picks up the even parity state, then Bob3 measures the photon
in the spatial mode $d_{3}$ in the basis $|\pm\rangle$. They will
get
\begin{eqnarray}
&&|\Psi^{\pm}\rangle^{3}_{N}=\frac{\alpha_{2}}{\sqrt{3\alpha^{2}_{2}+\alpha^{2}_{4}+\cdots\alpha^{2}_{N}}}|V\rangle_{1}|H\rangle_{2}|H\rangle_{3}|\widetilde{H}\rangle^{N-3}\nonumber\\
&&+\frac{\alpha_{2}}{\sqrt{3\alpha^{2}_{2}+\alpha^{2}_{4}+\cdots\alpha^{2}_{N}}}|H\rangle_{1}|V\rangle_{2}|H\rangle_{3}|H\rangle|\widetilde{H}\rangle^{N-3}\nonumber\\
&&\pm\frac{\alpha_{2}}{\sqrt{3\alpha^{2}_{2}+\alpha^{2}_{4}+\cdots\alpha^{2}_{N}}}|H\rangle_{1}|H\rangle_{2}|V\rangle_{3}|\widetilde{H}\rangle^{N-3}\nonumber\\
&&+\frac{\alpha_{4}}{\sqrt{3\alpha^{2}_{2}+\alpha^{2}_{4}+\cdots\alpha^{2}_{N}}}|H\rangle_{1}|H\rangle_{2}|H\rangle_{3}|V\rangle_{4}|\widetilde{H}\rangle^{N-4}\nonumber\\
&&+\cdots\nonumber\\
&&+\frac{\alpha_{N}}{\sqrt{3\alpha^{2}_{2}+\alpha^{2}_{4}+\cdots\alpha^{2}_{N}}}|H\rangle_{1}|H\rangle_{2}|H\rangle_{3}|\widetilde{H}\rangle^{N-4}|V\rangle_{N}.\label{W21}\nonumber\\
\end{eqnarray}
The total success probability is
\begin{eqnarray}
P^{3}=\frac{3\alpha^{2}_{2}\alpha^{2}_{3}+\alpha^{2}_{3}(\alpha^{2}_{4}+\cdots+\alpha^{2}_{N})}{(2\alpha^{2}_{2}+\alpha^{2}_{3}+\alpha^{2}_{4}+\cdots+\alpha^{2}_{N})(\alpha_{2}^{2}+\alpha_{3}^{2})}.
\end{eqnarray}
$P^{3}$ essentially contains two parts. The first one is the success
probability to get $|\Psi^{\pm}\rangle^{1}_{N}$, and the second one
is the success probability for Bob3 to pick up the even parity
state. Interestingly, from Eq. (\ref{W21}), the coefficient $\alpha_{3}$ has also
disappeared. The following concentration steps are similar to the above description. That is
each one performs a parity check measurement and picks up the even parity state. For instance, in the $Kth$ step, Bob$K$ first prepares
a single photon of the form
\begin{eqnarray}
|\Phi\rangle_{K}=\frac{\alpha_{2}}{\sqrt{\alpha_{2}^{2}+\alpha_{K}^{2}}}|V\rangle
+\frac{\alpha_{K}}{\sqrt{\alpha_{2}^{2}+\alpha_{K}^{2}}}|H\rangle.\label{SK}
\end{eqnarray}
After he performing the parity check measurement and picks up the even parity state, the original less-entangled $W$ state becomes
\begin{eqnarray}
&&|\Psi^{\pm}\rangle^{K}_{N}=\frac{\alpha_{2}}{\sqrt{K\alpha^{2}_{2}+\alpha^{2}_{K+1}+\cdots\alpha^{2}_{N}}}|V\rangle_{1}|\widetilde{H}\rangle^{N-1}\nonumber\\
&&+\frac{\alpha_{2}}{\sqrt{K\alpha^{2}_{2}+\alpha^{2}_{4}+\cdots\alpha^{2}_{N}}}|H\rangle_{1}|V\rangle_{2}|H\rangle_{3}|\widetilde{H}\rangle^{N-3}\nonumber\\
&&+\cdots\nonumber\\
&&\pm\frac{\alpha_{2}}{\sqrt{K\alpha^{2}_{2}+\alpha^{2}_{K+1}+\cdots\alpha^{2}_{N}}}|\widetilde{H}\rangle^{K-1}|V\rangle_{K}|\widetilde{H}\rangle^{N-K}\nonumber\\
&&+\frac{\alpha_{K+1}}{\sqrt{K\alpha^{2}_{2}+\alpha^{2}_{K+1}+\cdots\alpha^{2}_{N}}}|\widetilde{H}\rangle^{K}|V\rangle_{K+1}|\widetilde{H}\rangle^{N-K-1}\nonumber\\
&&+\cdots\nonumber\\
&&+\frac{\alpha_{N}}{\sqrt{K\alpha^{2}_{2}+\alpha^{2}_{K+1}+\cdots\alpha^{2}_{N}}}|\widetilde{H}\rangle^{N-1}|V\rangle_{N}.\label{WK1}
\end{eqnarray}
The success probability can be written as
\begin{eqnarray}
P^{K}=\frac{K\alpha^{2}_{2}\alpha^{2}_{K}+\alpha^{2}_{K}(\alpha^{2}_{K+1}+\alpha^{2}_{K+2}\cdots+\alpha^{2}_{N})}{((K-1)\alpha^{2}_{2}
+\alpha^{2}_{K}+\alpha^{2}_{K+1}+\cdots+\alpha^{2}_{N})(\alpha_{2}^{2}+\alpha_{K}^{2})}.\nonumber\\
\end{eqnarray}
If $K=N$, then they will get the maximally entangled $W$ state, with
the probability of
\begin{eqnarray}
P^{N}=\frac{N\alpha^{2}_{2}\alpha^{2}_{N}}{((N-1)\alpha^{2}_{2}+
\alpha^{2}_{N})(\alpha_{2}^{2}+\alpha_{N}^{2})}.
\end{eqnarray}

Therefore, the total success probability to get the maximally
entangled $W$ state from Eq. (\ref{partial W state}) is
\begin{eqnarray}
P_{T}=P^{1}P^{3}\cdots
P^{N}=\frac{N\alpha^{2}_{1}\alpha^{2}_{2}\alpha^{2}_{3}\cdots\alpha^{2}_{N}}
{(\alpha^{2}_{2}+\alpha^{2}_{1})(\alpha^{2}_{2}+\alpha^{2}_{3})\cdots(\alpha^{2}_{2}+\alpha^{2}_{N})}.\label{ProbabilityPPC}\nonumber\\
\end{eqnarray}
Interestingly, if $N=2$, it is the concentration of two-particle
Bell state with $P_{T}=2\alpha^{2}_{1}\alpha^{2}_{2}$. It is equal
to the success probability in Refs.\cite{zhao1,shengpra2,shengsinglephotonconcentration,dengsingle}.

By far, we have fully explained our ECP with PPC
gate. During the whole process, we require $N-1$ single photons to
achieve this task with the success probability of $P_{T}$. Except
Bob2, each parties needs to perform a parity check. If the parity check measurement result is even, it is
successful and he asks others to retain their photons. From
Sec. II,  the PPC gate essentially is based on linear optics and we should resort the post
selection principle. That is to say, the detection  will destroy
their photons. This disadvantage will greatly limits its practical
application,
because it has to require all of the parties to perform the parity
check simultaneously. On the other hand, the total success
probability is extremely low. Because they should ensure all $N-1$
parity checks  be successful. If any of parity check in Bob$K$ is
fail, then the whole ECP is fail. It is quite different from the ECP
of $N$-particle GHZ state\cite{zhao1,shengpra2,shengsinglephotonconcentration}, due to the
same entanglement structure with Bell state. The ECP of
Bell state is suitable to the $N$-particle GHZ state with the same
success probability $2\alpha^{2}_{1}\alpha^{2}_{2}$ with linear optics\cite{zhao1}. That is to say,
the success probability does not change with the particle number
$N$. However, in this ECP,
we find that the $P_{T}$ changes when $N$ changes.

\subsection{N-particle $W$ state concentration with CPC gate}
From above description, we show that the PPC gate can be used to
achieve this concentration task. However, it is not an economical
one and the success probability is extremely low. The reason is that we only pick up the even parity state and discard the
odd one. In this section, we will adopt the PPC gate to redescribe
this ECP. The basic principle of our ECP is shown in Fig. 4. We
use the CPC gates to substitute the PPC gates.
\begin{figure}[!h]%[tpb]
\begin{center}
\includegraphics[width=9cm,angle=0]{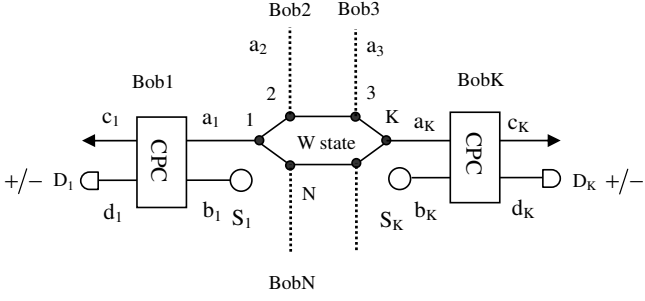}
\caption{A schematic drawing of our ECP with CPC
gate. Compared with Fig. 3, we use the CPC gate shown in Fig. 2 to substitute
the PPC gate. By using the CPC gate, the odd parity state can also be reused to improve the success probability and the concentrated state can also be retained. }
\end{center}
\end{figure}
In the first step, the initial state $|\Psi\rangle_{N}$  and $|\Phi\rangle_{1}$
combined with the coherent state $|\alpha\rangle$ evolve as
\begin{eqnarray}
&&|\Psi\rangle_{N+1}\otimes|\alpha\rangle=|\Psi\rangle_{N}\otimes|\Phi\rangle_{1}
=(\alpha_{1}|V\rangle_{1}|H\rangle_{2}|\widetilde{H}\rangle^{N-2}\nonumber\\
&+&\alpha_{2}|H\rangle_{1}|V\rangle_{2}|\widetilde{H}\rangle^{N-2}
+\cdots
+\alpha_{N}|H\rangle_{1}|H\rangle_{2}|\widetilde{H}\rangle^{N-3}|V\rangle_{N})\nonumber\\
&\otimes&(\frac{\alpha_{1}}{\sqrt{\alpha_{1}^{2}+\alpha_{2}^{2}}}|H\rangle
+\frac{\alpha_{2}}{\sqrt{\alpha_{1}^{2}+\alpha_{2}^{2}}}|V\rangle)\otimes|\alpha\rangle\nonumber\\
&\rightarrow&\frac{\alpha_{1}^{2}}{\sqrt{\alpha_{1}^{2}+\alpha_{2}^{2}}}|H\rangle|V\rangle_{1}|H\rangle_{2}|\widetilde{H}\rangle^{N-2}\otimes|\alpha e^{-i2\theta}\rangle\nonumber\\
&+&\frac{\alpha_{2}^{2}}{\sqrt{\alpha_{1}^{2}+\alpha_{2}^{2}}}|V\rangle|H\rangle_{1}|V\rangle_{2}|\widetilde{H}\rangle^{N-2}\otimes|\alpha e^{i2\theta}\rangle\nonumber\\
&+&\frac{\alpha_{1}\alpha_{2}}{\sqrt{\alpha_{1}^{2}+\alpha_{2}^{2}}}|H\rangle|H\rangle_{1}|V\rangle_{2}|\widetilde{H}\rangle^{N-2}\otimes|\alpha \rangle\nonumber\\
&+&\frac{\alpha_{1}\alpha_{3}}{\sqrt{\alpha_{1}^{2}+\alpha_{2}^{2}}}|H\rangle|H\rangle_{1}|H\rangle_{2}|V\rangle_{3}|\widetilde{H}\rangle^{N-3}\otimes|\alpha \rangle\nonumber\\
&+&\cdots\nonumber\\
&+&\frac{\alpha_{1}\alpha_{N}}{\sqrt{\alpha_{1}^{2}+\alpha_{2}^{2}}}|H\rangle|H\rangle_{1}|H\rangle_{2}|\widetilde{H}\rangle^{N-3}|V\rangle_{N}\otimes|\alpha \rangle\nonumber\\
&+&\frac{\alpha_{1}\alpha_{2}}{\sqrt{\alpha_{1}^{2}+\alpha_{2}^{2}}}|V\rangle|V\rangle_{1}|H\rangle_{2}|\widetilde{H}\rangle^{N-2}\otimes|\alpha \rangle\nonumber\\
&+&\frac{\alpha_{2}\alpha_{3}}{\sqrt{\alpha_{1}^{2}+\alpha_{2}^{2}}}|V\rangle|H\rangle_{1}|H\rangle_{2}|V\rangle_{3}|\widetilde{H}\rangle^{N-3}\otimes|\alpha e^{-i2\theta}\rangle \nonumber\\
&+&\cdots\nonumber\\
&+&\frac{\alpha_{2}\alpha_{N}}{\sqrt{\alpha_{1}^{2}+\alpha_{2}^{2}}}|V\rangle|H\rangle_{1}|H\rangle_{2}|\widetilde{H}\rangle^{N-3}|V\rangle_{N}\otimes|\alpha
e^{^{-i2\theta}} \rangle.
\end{eqnarray}

Obviously, if the coherent state $|\alpha\rangle$ picks up no phase
shift, the original state will collapse to the even
state similar to $|\Psi\rangle'_{N+1}$ in Eq. (\ref{collapse11}). It
can also be rewritten as $|\Psi\rangle''_{N+1}$ with the probability
of $P^{1}$. In this way, they can also obtain the same state
$|\Psi^{\pm}\rangle^{1}_{N}$ in Eq. (\ref{W11}) and can be used to
start the next concentration step on the number 3 photon performed by Bob3. On the
other hand, there is the probability of $1-P^{1}$ that the original
state will collapse to the odd state, if the phase shift of coherent
state is $2\theta$. Therefore, it can be written as

\begin{eqnarray}
|\Psi_{^{\bot}}\rangle'_{N+1}&=&\frac{\alpha_{1}^{2}}{\sqrt{\alpha_{1}^{2}+\alpha_{2}^{2}}}|H\rangle|V\rangle_{1}|H\rangle_{2}|\widetilde{H}\rangle^{N-2}\nonumber\\
&+&\frac{\alpha_{2}^{2}}{\sqrt{\alpha_{1}^{2}+\alpha_{2}^{2}}}|V\rangle|H\rangle_{1}|V\rangle_{2}|\widetilde{H}\rangle^{N-2}\nonumber\\
&+&\frac{\alpha_{2}\alpha_{3}}{\sqrt{\alpha_{1}^{2}+\alpha_{2}^{2}}}|V\rangle|H\rangle_{1}|H\rangle_{2}|V\rangle_{3}|\widetilde{H}\rangle^{N-3}\rangle \nonumber\\
&+&\cdots\nonumber\\
&+&\frac{\alpha_{2}\alpha_{N}}{\sqrt{\alpha_{1}^{2}+\alpha_{2}^{2}}}|V\rangle|H\rangle_{1}|H\rangle_{2}|\widetilde{H}\rangle^{N-3}|V\rangle_{N}.\nonumber\\
\end{eqnarray}
After measuring the photon in $d_{1}$ mode in the basis
$|\pm\rangle$, above state becomes
\begin{eqnarray}
|\Psi^{\pm}_{^{\bot}}\rangle'_{N}&=&\pm\frac{\alpha_{1}^{2}}{\sqrt{\alpha_{1}^{2}+\alpha_{2}^{2}}}|V\rangle_{1}|H\rangle_{2}|\widetilde{H}\rangle^{N-2}\nonumber\\
&+&\frac{\alpha_{2}^{2}}{\sqrt{\alpha_{1}^{2}+\alpha_{2}^{2}}}|H\rangle_{1}|V\rangle_{2}|\widetilde{H}\rangle^{N-2}\nonumber\\
&+&\frac{\alpha_{2}\alpha_{3}}{\sqrt{\alpha_{1}^{2}+\alpha_{2}^{2}}}|H\rangle_{1}|H\rangle_{2}|V\rangle_{3}|\widetilde{H}\rangle^{N-3}\rangle \nonumber\\
&+&\cdots\nonumber\\
&+&\frac{\alpha_{2}\alpha_{N}}{\sqrt{\alpha_{1}^{2}+\alpha_{2}^{2}}}|H\rangle_{1}|H\rangle_{2}|\widetilde{H}\rangle^{N-3}|V\rangle_{N}.\nonumber\\
\end{eqnarray}
If the measurement result is $|+\rangle$, they will get $|\Psi^{+}_{^{\bot}}\rangle'_{N}$, otherwise, they will get $|\Psi^{-}_{^{\bot}}\rangle'_{N}$.
Above equation can be rewritten as
\begin{eqnarray}
|\Psi^{\pm}_{^{\bot}}\rangle''_{N}&=&\pm\frac{\alpha_{1}^{2}}{T}|V\rangle_{1}|H\rangle_{2}|\widetilde{H}\rangle^{N-2}\nonumber\\
&+&\frac{\alpha_{2}^{2}}{T}|H\rangle_{1}|V\rangle_{2}|\widetilde{H}\rangle^{N-2}\nonumber\\
&+&\frac{\alpha_{2}\alpha_{3}}{T}|H\rangle_{1}|H\rangle_{2}|V\rangle_{3}|\widetilde{H}\rangle^{N-3}\rangle\nonumber\\
&+&\cdots\nonumber\\
&+&\frac{\alpha_{2}\alpha_{N}}{T}|H\rangle_{1}|H\rangle_{2}|\widetilde{H}\rangle^{N-3}|V\rangle_{N}.\label{lessentangled11}
\end{eqnarray}
$T=\sqrt{\alpha^{4}_{1}+\alpha^{2}_{2}(\alpha^{2}_{2}+\alpha^{2}_{3}+\cdots+\alpha^{2}_{N})}$.
 Interestingly, the state of  Eq. (\ref{lessentangled11}) essentially
is a lesser-entangled $W$ state. It can be reconcentrated with another
single photon on the number 1 photon. The another single photon is
written as

\begin{eqnarray}
|\Phi\rangle'_{1}=\frac{\alpha^{2}_{1}}{\sqrt{\alpha_{1}^{4}+\alpha_{2}^{4}}}|H\rangle
+\frac{\alpha^{2}_{2}}{\sqrt{\alpha_{1}^{4}+\alpha_{2}^{4}}}|V\rangle.\label{S11}
\end{eqnarray}
So Bob1 can restart this ECP with the help of a second single photon $|\Phi\rangle'_{1}$. The state $|\Psi^{+}_{^{\bot}}\rangle''_{N}$ and
$|\Phi\rangle'_{1}$ combined with the coherent state
$|\alpha\rangle$ evolves as
\begin{eqnarray}
&&|\Psi^{+}_{^{\bot}}\rangle''_{N}\otimes|\Phi\rangle'_{1}\otimes|\alpha\rangle=\frac{\alpha_{1}^{2}}{T}|V\rangle_{1}|H\rangle_{2}|\widetilde{H}\rangle^{N-2}\nonumber\\
&+&\frac{\alpha_{2}^{2}}{T}|H\rangle_{1}|V\rangle_{2}|\widetilde{H}\rangle^{N-2}\nonumber\\
&+&\frac{\alpha_{2}\alpha_{3}}{T}|H\rangle_{1}|H\rangle_{2}|V\rangle_{3}|\widetilde{H}\rangle^{N-3}\rangle\nonumber\\
&+&\cdots\nonumber\\
&+&\frac{\alpha_{2}\alpha_{N}}{T}|H\rangle_{1}|H\rangle_{2}|\widetilde{H}\rangle^{N-3}|V\rangle_{N}\nonumber\\
&\otimes&(\frac{\alpha^{2}_{1}}{\sqrt{\alpha_{1}^{4}+\alpha_{2}^{4}}}|H\rangle
+\frac{\alpha^{2}_{2}}{\sqrt{\alpha_{1}^{4}+\alpha_{2}^{4}}}|V\rangle)\otimes|\alpha\rangle\nonumber\\
&\rightarrow&\frac{\alpha^{4}_{1}}{T\sqrt{\alpha_{1}^{4}+\alpha_{2}^{4}}}|H\rangle|V\rangle_{1}|H\rangle_{2}|\widetilde{H}\rangle^{N-2}|\alpha e^{-i2\theta}\rangle\nonumber\\
&+&\frac{\alpha^{2}_{1}\alpha^{2}_{2}}{T\sqrt{\alpha_{1}^{4}+\alpha_{2}^{4}}}|V\rangle|V\rangle_{1}|H\rangle_{2}|\widetilde{H}\rangle^{N-2}|\alpha\rangle\nonumber\\
&+&\frac{\alpha^{2}_{1}\alpha^{2}_{2}}{T\sqrt{\alpha_{1}^{4}+\alpha_{2}^{4}}}|H\rangle|H\rangle_{1}|V\rangle_{2}|\widetilde{H}\rangle^{N-2}|\alpha\rangle\nonumber\\
&+&\frac{\alpha^{4}_{1}}{T\sqrt{\alpha_{1}^{4}+\alpha_{2}^{4}}}|V\rangle|H\rangle_{1}|V\rangle_{2}|\widetilde{H}\rangle^{N-2}|\alpha e^{i2\theta}\rangle\nonumber\\
&+&\frac{\alpha_{1}^{2}\alpha_{2}\alpha_{3}}{T\sqrt{\alpha_{1}^{4}+\alpha_{2}^{4}}}|H\rangle|H\rangle_{1}|H\rangle_{2}|V\rangle_{3}|\widetilde{H}\rangle^{N-3}\rangle|\alpha\rangle\nonumber\\
&+&\cdots\nonumber\\
&+&\frac{\alpha_{1}^{2}\alpha_{2}\alpha_{N}}{T\sqrt{\alpha_{1}^{4}+\alpha_{2}^{4}}}|H\rangle|H\rangle_{1}|H\rangle_{2}|\widetilde{H}\rangle^{N-3}|V\rangle_{N}|\alpha\rangle\nonumber\\
&+&\frac{\alpha_{2}^{3}\alpha_{3}\alpha_{N}}{T\sqrt{\alpha_{1}^{4}+\alpha_{2}^{4}}}|V\rangle|H\rangle_{1}|H\rangle_{2}|V\rangle_{3}|\widetilde{H}\rangle^{N-3}|\alpha e^{i2\theta}\rangle\nonumber\\
&+&\cdots\nonumber\\
&+&\frac{\alpha_{2}^{3}\alpha_{N}\alpha_{N}}{T\sqrt{\alpha_{1}^{4}+\alpha_{2}^{4}}}|V\rangle|H\rangle_{1}|H\rangle_{2}|\widetilde{H}\rangle^{N-3}|V\rangle_{N}|\alpha e^{i2\theta}\rangle\nonumber\\
\end{eqnarray}

\begin{figure}[!h]%[tpb]
\begin{center}
\includegraphics[width=7cm,angle=0]{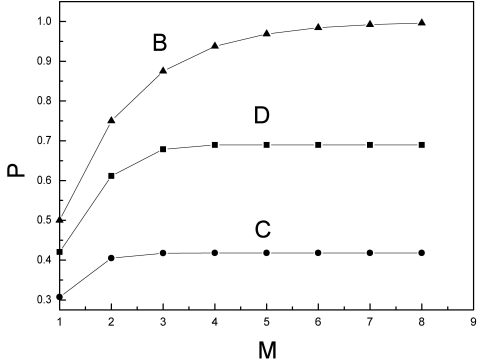}
\caption{The success probability of concentration of each photon in
each step is altered with the iteration number $M$. Here we choose
the five-photon less entangled $W$ state with
$\alpha_{1}=\alpha_{2}=\alpha_{3}=0.5$, $\alpha_{4}=0.3$,
$\alpha_{5}=0.4$. Curve B: the success probability of concentration
the number 1 and 3 photons according to $\alpha_{1}=\alpha_{3}=0.5$.
Curve C: the success probability of concentration the number 4
photon according to $\alpha_{4}=0.3$. Curve D: the success
probability of concentration the number 4 photon according to
$\alpha_{5}=0.4$.}
\end{center}
\end{figure}
Obviously, if Bob1 picks up no phase shift, above equation
collapses to
\begin{eqnarray}
&&|\Psi_\perp\rangle'_{N+1}=\frac{\alpha^{2}_{1}\alpha^{2}_{2}}{T\sqrt{\alpha_{1}^{4}+\alpha_{2}^{4}}}|H\rangle|H\rangle_{1}|V\rangle_{2}|\widetilde{H}\rangle^{N-2}\nonumber\\
&+&\frac{\alpha^{2}_{1}\alpha^{2}_{2}}{T\sqrt{\alpha_{1}^{4}+\alpha_{2}^{4}}}|H\rangle|H\rangle_{1}|V\rangle_{2}|\widetilde{H}\rangle^{N-2}\nonumber\\
&+&\frac{\alpha_{1}^{2}\alpha_{2}\alpha_{3}}{T\sqrt{\alpha_{1}^{4}+\alpha_{2}^{4}}}|H\rangle|H\rangle_{1}|H\rangle_{2}|V\rangle_{3}|\widetilde{H}\rangle^{N-3}\rangle|\alpha\rangle\nonumber\\
&+&\cdots\nonumber\\
&+&\frac{\alpha_{1}^{2}\alpha_{2}\alpha_{N}}{T\sqrt{\alpha_{1}^{4}+\alpha_{2}^{4}}}|H\rangle|H\rangle_{1}|H\rangle_{2}|\widetilde{H}\rangle^{N-3}|V\rangle_{N}.
\end{eqnarray}
\begin{figure}[!h]%[tpb]
\begin{center}
\includegraphics[width=7cm,angle=0]{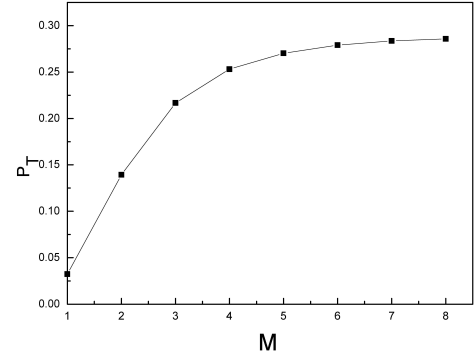}
\caption{The total success probability of our ECP
altered with iteration number $M$ with CPC gate. We also let
$\alpha_{1}=\alpha_{2}=\alpha_{3}=0.5$, $\alpha_{4}=0.3$,
$\alpha_{5}=0.4$.}
\end{center}
\end{figure}
Interestingly, above state essentially is the state
$|\Psi\rangle''_{N+1}$ in Eq.(\ref{rewrittecollapse11}), if it is
normalized. Then it can be used to concentrate the number 3 photon
with the same CPC gate like above. The success probability is
\begin{eqnarray}
P_{2}^{1}=\frac{2\alpha^{4}_{1}\alpha^{4}_{2}+\alpha^{4}_{1}\alpha^{2}_{2}(\alpha^{2}_{3}+\alpha^{2}_{4}+\cdots+\alpha^{2}_{N})}
{(\alpha^{2}_{1}+\alpha^{2}_{2})(\alpha^{4}_{1}+\alpha^{4}_{2})}.
\end{eqnarray}
Following the same principle, Bob1 can repeat this ECP for $M$ times and they can  get the success probability in each step as
\begin{eqnarray}
P_{3}^{1}&=&\frac{2\alpha^{8}_{1}\alpha^{8}_{2}+\alpha^{8}_{1}\alpha^{6}_{2}(\alpha^{2}_{3}+\alpha^{2}_{4}+\cdots+\alpha^{2}_{N})}
{(\alpha^{2}_{1}+\alpha^{2}_{2})(\alpha^{4}_{1}+\alpha^{4}_{2})(\alpha^{8}_{1}+\alpha^{8}_{2})},\nonumber\\
&\cdots&\nonumber\\
P_{M}^{1}&=&\frac{2\alpha^{2^{M}}_{1}\alpha^{2^{M}}_{2}+\alpha^{2^{M}}_{1}\alpha^{2^{M}-2}_{2}(\alpha^{2}_{3}+\alpha^{2}_{4}+\cdots+\alpha^{2}_{N})}
{(\alpha^{2}_{1}+\alpha^{2}_{2})(\alpha^{2^{2}}_{1}+\alpha^{2^{2}}_{2})\cdots(\alpha^{2^{M}}_{1}+\alpha^{2^{M}}_{2})}.\nonumber\\
\end{eqnarray}
Here the superscription 1 means that concentration on the number 1
photon. The subscription $M$ means that the ECP is performed $M$
times.

After performing the concentration ECP on the number 1 photon, they
will have a total success probability with
$P_{total}^{1}=P^{1}_{1}+P^{1}_{2}+\cdots+P^{1}_{M}=\sum^{\infty}_{M=1}
P^{1}_{M}$ to obtain $|\Psi^{\pm}\rangle^{1}_{N}$, which can be used to
performing the concentration scheme on the number 3 photon.
So far, we have explained our ECP performed on the
number 1 photon with CPC gate. Different from the scheme described
with PPC gate, it can be repeated to get a high success probability.

Following the same principle, they can also use this way to
concentrating each photons. If they perform this ECP
on the $Kth$ ($K\neq 2$) photon with $M$ times, they can get the success
probability $P^{K}_{M}$
\begin{widetext}
\begin{eqnarray}
P^{K}_{M}=\frac{K\alpha^{2^{M}}_{2}\alpha^{2^{M}}_{K}+\alpha^{2^{M}}_{K}\alpha^{2^{M}-2}_{2}(\alpha^{2}_{K+1}+\alpha^{2}_{K+2}+\cdots+\alpha^{2}_{N})}
{[(K-1)\alpha^{2}_{2}+\alpha^{2}_{K}+\alpha^{2}_{K+1}+\cdots\alpha^{2}_{N}][(\alpha^{2}_{2}+\alpha^{2}_{K})(\alpha^{2^{2}}_{2}+\alpha^{2^{2}}_{K})\cdots(\alpha^{2^{M}}_{2}+\alpha^{2^{M}}_{K})]}.
\end{eqnarray}
\end{widetext}
If $K=N$, they can get
\begin{widetext}
\begin{eqnarray}
P^{N}_{M}=\frac{N\alpha^{2^{M}}_{2}\alpha^{2^{M}}_{N}}
{[(N-1)\alpha^{2}_{2}+\alpha^{2}_{N}][(\alpha^{2}_{2}+\alpha^{2}_{N})(\alpha^{2^{2}}_{2}+\alpha^{2^{2}}_{N})\cdots(\alpha^{2^{M}}_{2}+\alpha^{2^{M}}_{N})]}.\label{Pmn}
\end{eqnarray}
\end{widetext}
Therefore, if we use the CPC gate to perform the EPC, each parties can repeat
this ECP to increase the success probability. Suppose each one all perform this ECP
for $M$ times, the success probability of get a maximally entangled $W$ state
from the initial state in Eq. (\ref{partial W state}) can be
described as
\begin{eqnarray}
P&=&P^{1}_{total}P^{3}_{total}P^{4}_{total}\cdots P^{N}_{total}\nonumber\\
&=&(P^{1}_{1}+P^{1}_{2}+\cdots+P^{1}_{M})(P^{3}_{1}+P^{3}_{2}\nonumber\\
&+&\cdots+P^{3}_{M})\cdots(P^{N}_{1}+P^{N}_{2}+\cdots+P^{N}_{M})\nonumber\\
&=&\prod^{N}_{K=1,K\neq2}(\sum^{\infty}_{M=1}
P^{K}_{M}).\label{ProbabilityCPC}
\end{eqnarray}
Compared with the ECP with PPC gate, the success probability in
Eq. (\ref{ProbabilityPPC}) is the case of $M=1$ in
Eq. (\ref{ProbabilityCPC}).

In Fig. 5, we show that the success probability of concentration of
each photon altered with the iteration number $M$. We take the
five-photon less-entangled $W$ state as an example. We  let
$\alpha_{1}=\alpha_{2}=\alpha_{3}=0.5$, $\alpha_{4}=0.3$ and
$\alpha_{5}=0.4$. Interestingly, if
$\alpha_{1}=\alpha_{2}=\alpha_{3}=0.5$, the success probability of
concentration number 1 photon $P_{M}^{1}$ is equal to $P_{M}^{3}$,
shown in Curve B. We calculated the total success probability of our
protocol with CPC gate shown in Fig.6. It is shown that, if we use the
PPC gate, the success probability is the case of $M=1$, that is
0.03228. But if we use the CPC gate and iterate it for eight times,
the success probability can be increased to 0.28575. It is about
nine times greater than the success probability of using PPC gate.

\section{discussion and summary}
So far, we have fully described our ECP for
N-particle less-entangled $W$ state. We explain this ECP with two different
methods. The first one is to use the PPC gates  and the second one is to use the CPC gates. In our ECP,
after successfully performing this parity check, all coefficients in the
initial state are equal to $\alpha_{2}$. In fact, this is not the
unique way to achieve this task. We can also choose $\alpha_{1}$ and  make all coefficients be equal to $\alpha_{1}$ after performing this ECP. Choosing  different coefficients do not change the basic principle of this ECP, but it will change
the total success probability.
In detail, we take
four-particle less-entangled $W$ state and  five-particle less-entangled  $W$ state for example. Fig. 7 shows the success probability altering
with the iteration number $M$ for
case of four-particle.
\begin{figure}[!h]%[tpb]
\begin{center}
\includegraphics[width=7cm,angle=0]{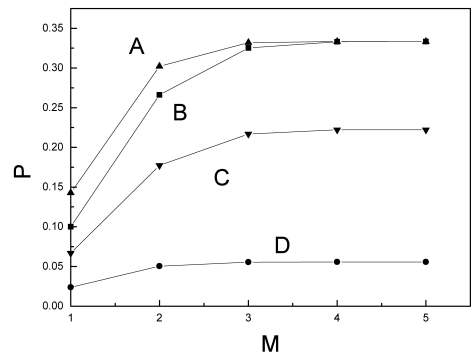}
\caption{The total success probability of our ECP
for four-partite $W$ state altered with iteration number $M$ with CPC
gate. Curve A: $\alpha_{1}=\frac{1}{\sqrt{6}}$,
$\alpha_{2}=\frac{1}{\sqrt{12}}$, $\alpha_{3}=\frac{1}{\sqrt{2}}$,
$\alpha_{4}=\frac{1}{2}$. Curve B: $\alpha_{1}=\frac{1}{2}$,
$\alpha_{2}=\frac{1}{\sqrt{6}}$, $\alpha_{3}=\frac{1}{\sqrt{2}}$,
$\alpha_{4}=\frac{1}{2}$. Curve C: $\alpha_{1}=\frac{1}{\sqrt{2}}$,
$\alpha_{2}=\frac{1}{2}$, $\alpha_{3}=\frac{1}{\sqrt{6}}$,
$\alpha_{4}=\frac{1}{\sqrt{12}}$. Curve D:
$\alpha_{1}=\frac{1}{\sqrt{12}}$, $\alpha_{2}=\frac{1}{\sqrt{2}}$,
$\alpha_{3}=\frac{1}{2}$, $\alpha_{1}=\frac{1}{\sqrt{6}}$. }
\end{center}
\end{figure}
In Fig. 7, the less-entangled $W$ states corresponding to different
curves essentially have the same entanglement. Because they can change to each
other with local operations. However,  it is shown that
the same initial  entanglement  have the different success
probabilities if we choose different $\alpha_{2}$. In Fig. 8, we also
calculate the  similar case of five-particle less-entangled $W$ state. One can see that
choosing different $\alpha_{2}$ leads different total success
probability. That is $\alpha_{2}$ smaller, the total success
probability is greater.
\begin{figure}[!h]%[tpb]
\begin{center}
\includegraphics[width=7cm,angle=0]{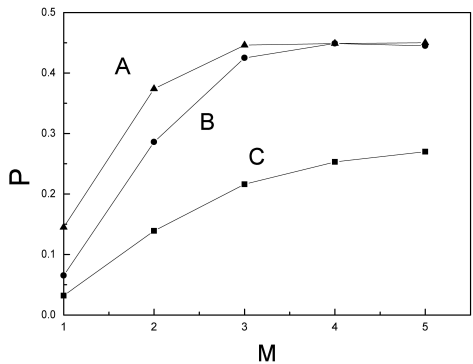}
\caption{The total success probability of our ECP
for five-particle $W$ state altered with iteration number $M$ with CPC
gate. Curve A: $\alpha_{1}=0.4$, $\alpha_{2}=0.3$,
$\alpha_{3}=\alpha_{4}=\alpha_{5}=0.5$. Curve B: $\alpha_{1}=0.5$,
$\alpha_{2}=0.4$, $\alpha_{3}=0.3$, $\alpha_{4}=\alpha_{5}=0.5$.
Curve C: $\alpha_{1}=\alpha_{2}=\alpha_{3}=0.5$, $\alpha_{4}=0.3$,
$\alpha_{5}=0.4$.}
\end{center}
\end{figure}
We can  explain this result from Eq. (\ref{ProbabilityPPC}) and
Eq. (\ref{Pmn}).  In Eq. (\ref{ProbabilityPPC}), if $\alpha_{1},
\alpha_{2},\cdots \alpha_{N}$ are given, the numerator is a
constant. But the value of denominator is decided by $\alpha_{2}$.
Therefore, choosing the smallest value of $\alpha_{2}$ will get the
highest  success probability. This result provide us an
useful way to performing this ECP. If $\alpha_{2}$
is not the smallest one, then we can rotate the polarization of each
photon with half-wave plate until to obtain the smallest $\alpha_{2}$ .
Certainly,  we should point out that the ECP with PPC gate and with CPC gate
 are quite different form each
other in the practical manipulation. Because the  PPC gate is equipped with the
linear optical elements and we should resort the pose selection
principle to achieve this task. That is, after successfully performed this ECP, the maximally entangled $W$
state is destroyed by the sophisticated single photon detectors. This condition greatly limit
its practical application. In addition,  in Sec. III A, we  explain it by
dividing the whole ECP into $N-1$ steps. In each step, one of the parties  prepares one single photon
and makes a parity check measurement.
Practically, each parties except Bob2 should perform the parity
check measurement simultaneously due to the post selection
principle. If all  parity check measurements are even parities,
then by classical communication, they ask each to retain their
photons, and it is a successful case. On the other hand, if we
adopt CPC gate to perform this ECP, each parties can operate his
photons independently. That is, each one can repeat to perform
concentration until it is successful. The most fundamental reason is
that QND is only to check the phase shift of the coherent state and it does not destroy the photon after measurement. In our
ECP, after performing the parity check measurement using CPC
gate, the next operation is decided by the measurement result. If it
is even parity, it is successful, otherwise, each one can restart to
concentrate his photon with another single photon. This strategy
makes the total concentration efficiency be greatly improved.

Finally, let us discuss the key element of our ECP, that is the
cross-Kerr nonlinearity. In Ref. \cite{shengpra2} and
 \cite{shengqic}, they also adopt the cross-Kerr nonlinearity to
construct the parity check gate to achieve the concentration tasks.
Unfortunately, in order to increase the efficiency of the protocol,
they should resort the coherent state to obtain $\pi$ phase shift.
Although there are several strategies to increase the phase shift,
such as increasing the strength of the coherent state, controlling
the coupling  time of the coherent state and the Kerr media, and
choosing the suitable Kerr media, to get giant phase shift is still
difficult in current technology \cite{kok1,kok2}. Meanwhile,
cross-Kerr nonlinearity is also a controversial topic. The focus of
the argument is still that one cannot get giant phase shift on the
single-photon level. This conclusion is agree with the results of
Shapiro,  Razavi, and Gea-Banacloche\cite{Gea,Shapiro1,Shapiro2}. On
the other hand, Hofmann pointed out that with a single two-level
atom in a one-sided cavity, a large phase-shift of $\pi$ can be
achieved\cite{hofmann}. Current research showed that it is
possible to amplify a cross-Kerr-phase-shift to an observable
value by using weak measurements, which is much larger than the intrinsic magnitude of the
single-photon-level nonlinearity\cite{weak_meaurement}.  Zhu and Huang also discussed
the possibility of obtain the giant cross-Kerr nonlinearities using a double-quantum-well structure
with a four-lever, double-type configuration\cite{oe}.
Fortunately, we do not require the coherent
state to get $\pi$ phase shift. It is an improvement of
Refs. \cite{shengpra2,shengqic,shengsinglephotonconcentration}. This
kind of parity check gate is first used to perform the entanglement
purification in Ref. \cite{shengpra}. Then Guo \emph{et al.}
developed this idea, and used it to perform the Bell-state analyzer,
prepare the cluster-state and so on \cite{qi}. As discussed by Guo
\emph{et al.}, compared with the previous parity check gate\cite{shengpra2,shengqic,shengsinglephotonconcentration}, it has
several advantages: first, it is an effective simplification by
removing two PBSs and several mirrors; second, it has a lower error
rate. Third, it does not require the $\pi$ phase shift which is more
suitable in current experimental conditions.

In summary, we have present an universal way to concentrate an
N-particle less-entangled $W$ state into a maximally entangled $W$ state
with both PPC gate and CPC gate. In the former, we require the
linear optical elements and post selection principle.  In the later,
we use cross-Kerr nonlinearity to construct the QND.  Different from
other concentration protocols, we only need single photon as an
auxiliary to achieve the task. Then this ECP does not largely consume the
less-entangled photon systems. Especially, with the help of QND,
each parties can operated independently and  this ECP can be
repeated to get a higher success probability.  These advantages
may make this ECP more useful in practical application in current
quantum information processing.

\section*{ACKNOWLEDGEMENTS}
This work is supported by the National Natural Science Foundation of
China under Grant No. 11104159, Scientific Research
Foundation of Nanjing University of Posts and Telecommunications
under Grant No. NY211008,  University Natural Science Research
Foundation of JiangSu Province under Grant No. 11KJA510002, and the
open research fund of Key Lab of Broadband Wireless Communication
and Sensor Network Technology (Nanjing University of Posts and
Telecommunications), Ministry of Education, China, and the Project
Funded by the Priority Academic Program Development of Jiangsu
Higher Education Institutions.

\end{document}